\documentclass[3p,times,twocolumn]{elsarticle}

\usepackage{ecrc}

\volume{00}

\firstpage{1}

\journalname{Nuclear Physics B Proceedings Supplement}

\runauth{}

\jid{nuphbp}

\jnltitlelogo{Nuclear Physics B Proceedings Supplement}

\usepackage{amssymb}

\usepackage[figuresright]{rotating}

\begin{document}

\begin{frontmatter}



\dochead{}

\title{ IGR J17354$-$3255 as  bench test for investigation of $\gamma$-ray emission from Supergiant Fast X-ray Transients}


\author{Sguera V.}

\address{INAF/IASF Bologna, Via Piero Gobetti 101, Bologna}

\begin{abstract}

Among the different types of  sources shining in the high energy sky,  gamma-ray binaries  are rapidly becoming the subject of major interest. In fact,   in the last  few years  a number of High Mass X-ray Binaries (HMXBs) have been firmly detected  from MeV to TeV energies, providing secure evidences that particles can be efficiently accelerated  up to very high energies in such galactic systems. Similarly to this general and emerging class of gamma-ray binaries,  in principle  Supergiant Fast X-ray Transients (SFXTs) have all the "ingredients" to be  
transient high energy emitters. In this context, the SFXT  IGR J17354$-$3255 is a good bench test and we
present intriguing hints likely suggesting that  it is a transient gamma-ray source flaring on short timescales.
If fully confirmed by further studies,  the implications  stemming are huge, both theoretically and observationally, and would 
add a further extreme characteristic  to the already extreme class of SFXTs. 

\end{abstract}

\begin{keyword}


\end{keyword}

\end{frontmatter}


\section{Introduction}
\label{}
During the last few years,  AGILE and Fermi observations of the Galactic plane have indicated the existence of a 
possible population of unidentified transient MeV-GeV sources characterized by flares lasting no more than a few days at most 
e.g. \cite{hays}, \cite{bulgarelli}. Notably, no blazar-like counterparts are known  within their error boxes so 
they could represent a completely new class of Galactic high energy transients. The task of identifying their counterparts at lower 
energies is very challenging, mainly because of their fast transient nature and large positional uncertainty  (e.g radii typically 
from 10 arcmin  to 0.5 degrees). However these difficulties are compensate by the fact that it is most probably within 
this group of gamma-ray sources that peculiar objects (or even a new class of objects) could emerge,  leading  to novel and unexpected discoveries.  The IBIS instrument onboard the INTEGRAL satellite is particularly suited to search for reliable   best candidate counterparts in the energy range 20--100 keV  thanks to i) a large field of view  (30$^{\circ}$$\times$30$^{\circ}$) which ensure a total coverage of the gamma-ray error box  ii) a good angular resolution (12 arcminutes) which is mandatory to disentangle the hard X-ray emission of different 
sources in crowded fields such as those on the Galactic plane iii) a good sensitivity above 20 keV ($\sim$ 10 mCrab during a typical IBIS observation lasting $\sim$ 2,000 seconds).  In particular, recent INTEGRAL/IBIS results \cite{sguera1}, \cite{sguera2}, \cite{sguera3}  provided intriguing hints that best  candidate counterparts could be found among the members of the
 Supergiant Fast X-ray Transients population  (SFXTs). 

SFXTs are a new subclass of High Mass X-ray Binaries (HMXBs)  unveiled in the last few years mainly thanks to INTEGRAL observations of 
the Galactic plane \cite{sguera4}, \cite{sguera5}, \cite{negueruela}. They host a massive OB supergiant star as identified by optical spectroscopy,   the compact object is generally assumed to be a neutron star because of the broad band X-ray spectral shape  (0.2--100 keV) strongly resembling  those of accreting X-ray pulsars in classical HMXBs. As support to this assumption, X-ray pulsations have been firmly detected in several  SFXTs e.g. \cite{sguera6}.  As for their X-ray behavior, SFXTs are intriguingly characterized by remarkable fast  X-ray flares lasting from few hours to no longer than a few days 
at most and reaching  typical peak X-ray luminosities of  L$_{x}$ $\sim$ 10$^{36}$ erg s$^{-1}$ . 
The duty cycles of activity are very low (0.1\%--3\%), conversely  SFXTs  spend most of their time in a low level X-ray activity 
with typical  L$_x$ $\sim$ 10$^{33}$--10$^{34}$ erg s$^{-1}$,  rarely  they are also observed in  X-ray quiescence 
at much lower X-ray luminosities (L$_x$ $\le$ 10$^{32}$ erg s$^{-1}$). 
The typical  dynamic range of classical SFXT spans three to five order of magnitude,  however some systems 
show a lower value of the order of $\sim$10$^{2}$ and so they have been named as intermediate SFXTs \cite{clark}, \cite{wz}. The extreme case of fast variability and high dynamical range characterizing SFXTs  is at odds with the behavior of their historical parent population of classical wind-fed supergiant HXMBs which, since more than 40 years of X-ray astronomy,  are known to be bright and persistent  X-ray sources always detectable around  L$_{x}$ $\sim$ 10$^{36}$ erg s$^{-1}$ . 
Although SFXT hunting is not an easy task, in a very few years $\sim$ 10 firm systems have been reported in the literature (see list in \cite{grebenev}) plus a similar number of candidates: SFXTs could represent a major population of  transient  HMXBs hidden on the Galactic plane of our Galaxy. The physical mechanism driving their  peculiar fast X-ray transient behaviour  is unclear and still highly debated.  Several models  have been proposed in the literature (see \cite{sidoli1} for a review). 

It is worth noting that in the last few years observations performed by Fermi, AGILE and Cherenkov telescopes have provided secure evidences that particles can be efficiently accelerated  to very high energies in some HMXBs. In fact, a number of classical HMXBs  have been firmly detected from MeV to TeV energies as persistent and variable sources e.g.  \cite{mirabel}. In addition, the two microquasar HMXB
Cygnus X-1 and Cygnus X-3 have been detected as transient MeV-GeV  emitters whose flaring activity lasted typically 1-2 days 
\cite{tavani}, \cite{abdo}, \cite{sabatini}.  Similarly to this general and emerging class of gamma-ray HMXBs, in principle SFXTs 
could  be able to produce photons up to MeV-TeV energies since they have the  same "ingredients" in term of  a neutron star compact object as well as a bright and massive OB star which could act as a source of seed photons (for the Inverse Compton emission) and target nuclei (for hadronic interactions). In particular, the  high energy  emission from SFXTs should be in the form of fast flares (from few hours to few days duration) and the duty cycle of activity should be very small,  this would make  their high energy detection very challenging. 
Despite this drawback, some observational evidences have been recently  reported in the literature on SFXTs
as best candidate counterparts of unidentified transient MeV-GeV sources located on the Galactic plane \cite{sguera1}, \cite{sguera2}, \cite{sguera3}. 
These evidences are merely based on intriguing hints such as a spatial correlation and  
a common transient behaviour on  similar, though as yet not simultaneous, short time scales. 
This scenario is also supported from an energetic  standpoint by a theoretical model based in the microquasar 
accretion/jet  framework \cite{sguera1}. The so far proposed associations represent an important first step towards 
obtaining reliable candidates on which to concentrate further efforts in order to obtain quantitative proofs for 
a real physical association.  In this respect, so far, the best test case is represented by the proposed association
between the two sources IGR J17354$-$3255 and AGL J1734$-$3310.

\section{The SFXT IGR J17354$-$3255}

\begin{figure}
\begin{center}
\includegraphics[width=.5\textwidth]{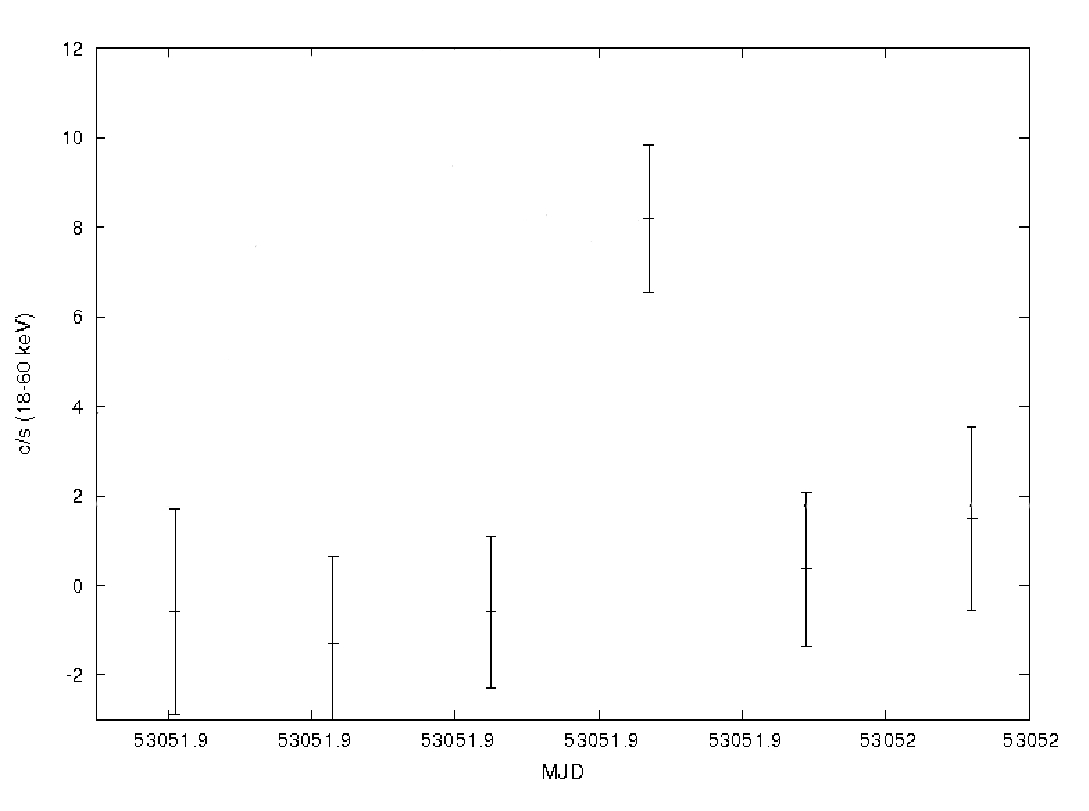}
\caption{The INTEGRAL/IBIS  light curve (18--60 keV) of a flare from  IGR J17354$-$3255. Each data point represents the average flux during 
one pointing (ScW) lasting 2,000 seconds.} 
\end{center}
\end{figure}

\begin{figure*}
\begin{center}
\includegraphics[width=.66\textwidth]{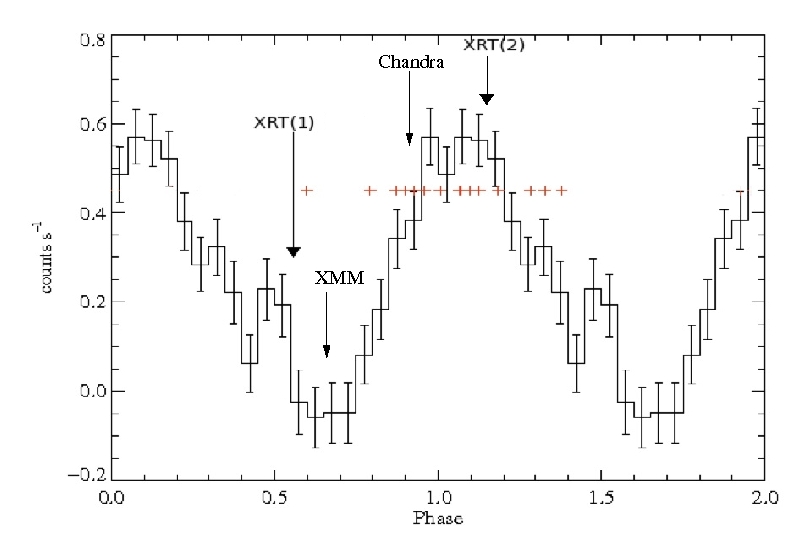}
\caption{ Phase-folded light curve of IGR J17354$-$3255. We note that the feature at phase  about 0.5 is due to the binning and is not taken as being of physical origin. The red crosses mark all the outbursts detected by IBIS. We marked by arrows the soft X-ray observations 
during which the source was not detected (XRT1 and XMM) as well as the ones during which it  was detected (XRT2 and Chandra).} 
\end{center}
\end{figure*}

IGR J17354$-$3255 is a hard X-ray transient discovered by INTEGRAL in 2006 during 
an outburst having average flux of $\sim$ 2.1$\times$10$^{-10}$ erg cm$^{-2}$ s$^{-1}$
(20--60 keV) and unconstrained duration (\cite{k1}, \cite{k2}. Subsequent studies  with the Swift satellite  
showed that its broad band X-ray spectrum (0.2--100 keV) is characterized by a spectral shape very similar to that of  
accreting X-ray pulsars in HMXBs, in addition  a periodicity of  $\sim$ 8.4 days  was unveiled in the Swift/BAT data \cite{dai}.  
In the soft X-ray domain (0.2--10 keV), the source has been  observed by Swift/XRT \cite{ver},
Chandra \cite{tom} and XMM \cite{boz}. Specifically, during  two observations (March 2008 with XRT and March 2011 with XMM) the source was not detected leading to an inferred  3$\sigma$  upper limit of 7$\times$10$^{-14}$ erg cm$^{-2}$ s$^{-1}$, conversely during  two other observations  (February 2009 with Chandra and  April 2009 with XRT) the source was indeed detected and it was strongly variable as well, being the  average flux  and the peak flux equal to $\sim$ 1.3$\times$10$^{-11}$ erg cm$^{-2}$ s$^{-1}$ and $\sim$ 9$\times$10$^{-11}$ erg cm$^{-2}$ s$^{-1}$, respectively. In addition, Chandra  provided a very accurate positioning which allowed to pinpoint a single  bright 2MASS infrared source as counterpart \cite{tom}.  Subsequent infrared spectroscopic performed by \cite{coleiro} unveiled its nature of supergiant star with  spectral type O9.5Iab. 

The temporal X-ray behavior of IGR J17354$-$3255 above 20 keV, which is crucial to allow a proper and firm classification as classical persistent supergiant HMXB or alternatively as a SFXT, is largely unknown. We performed a detailed temporal study with INTEGRAL in the energy band 18--60 keV  \cite{sguera2}. As result, we found that IGR J17354$-$3255 is a weak persistent  hard X-ray source spending  a major fraction of the time in  a out-of-outburst state with an average 18--60 keV flux of  $\sim$ 1.4$\times$10$^{-11}$ erg cm$^{-2}$ s$^{-1}$. This is occasionally interspersed with fast hard X-ray flares having  duration in the range  0.5--60 hours, a total of 16 hard X-ray flares have been detected by INTEGRAL/IBIS  over a total exposure of $\sim$ 115 days though not in sequence. 
Fig. 1 shows an example of a fast flare lasting only half an hour and detected with a  significance of about 6 $\sigma$ (18--60 keV). 
The dynamic range of the source in the hard X-ray band  (18--60 keV) is  as high as 200,  it is 
even higher ($>$1,250)  in the  softer X-ray  band  (0.2--10 keV).   Based on the above findings, IGR J17354$-$3255 can be classified as 
a member of the SFXT population.  In the framework of the clumpy wind scenario,  the fast X-ray flares 
could be explained as due to accretion onto the compact object of dense clumps of material in the highly structured wind of the supergiant companion donor star. Conversely, during the   weak persistent out-of-outburst X-ray state, accretion is likely still at work although at much lower rate, e.g. the compact object is likely accreting from the much less dense background wind. This  is supported by the fact that the 18--60 keV spectral shape, as measured by INTEGRAL/IBIS,  is identical during both the flaring and  out-of-outburst X-ray states, i.e. a power law shape with $\Gamma$=2.4$\pm$0.4.  

\begin{figure}
\includegraphics[width=.5\textwidth]{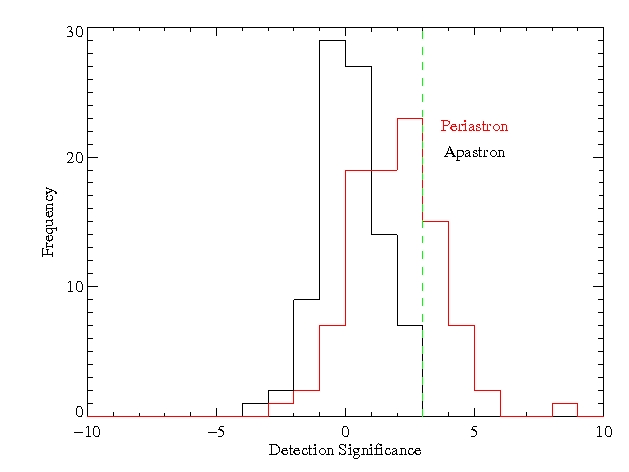}
\caption{Recurrence analysis for periastron and apastron detections of IGR J17354$-$3255 within a set window of 3 days.} 
\end{figure}

Our investigation of  the INTEGRAL/IBIS  long-term light curve using the Lomb-Scargle periodogram method  strongly confirmed  the 8.4 days  orbital period. Fig. 2 shows the corresponding phase folded light curve where it is evident a smooth orbital modulation of the flux; it peaks during the periastron passage and becomes consistent with zero around apastron. The red crosses indicate the 16 outbursts detected by INTEGRAL/IBIS within these orbital ephemeris. As clearly evident their occurrence is consistent with the region of orbital phase around periastron. We note that the shape of the orbital profile  is rather smooth and appears to be dominated by lower level X-ray emission rather than by the bright X-ray outbursts. To test this assumption we employed the recurrence analysis technique which searches for periastron detections  by summing the X-ray emission during each periastron passage within a set window of 3 days (periastron $\pm$ 1.5 days). The source could then be detected even though a significant detection is not achieved in the individual IBIS pointing lasting 2,000 seconds.  The same process was performed for each apastron passage and the distributions compared. Fig. 3 shows the recurrence analysis results, there is a clear excess in detections above 3$\sigma$ during the periastron passages, corresponding to detectable emission on about 26\% of periastron passages covered by the data set. This value is taken as a lower limit as there may still be emission that is occurring during other periastron passages that is below the sensitivity of IBIS. On the contrary no detections above 3$\sigma$ are recorded during apastron passages. Our recurrence analysis suggests that the 16 individual outbursts detected by IBIS/ISGRI cannot explain the smooth shape seen in the phase folded light curve. In fact,  assuming a source distance of 8.5 kpc (the distance is still unknown and the source is located in the direction of the Galactic center) these outbursts all have X-ray luminosities of the order of 10$^{36}$ erg s$^{-1}$ and so represent the most luminous X-ray outburst events. Hence we would not expect these events to define the orbital emission profile over the extent of these long baseline observations. Instead we attribute the shape to lower level X-ray emission that is below the instrumental sensitivity of IBIS in an individual ScW (i.e. $\sim$ 10 mCrab). However when the whole data set, covering about 300 orbital cycles of 8.4 days, is folded this emission sums to a significant detection and reveals the smooth profile shown in Fig. 2. This emission could be the superposition of many lower intensity X-ray flares at luminosity values of  $\sim$ 10$^{33}$--10$^{34}$ erg s$^{-1}$.

\begin{figure*}
\begin{center}
\includegraphics[width=.9\textwidth]{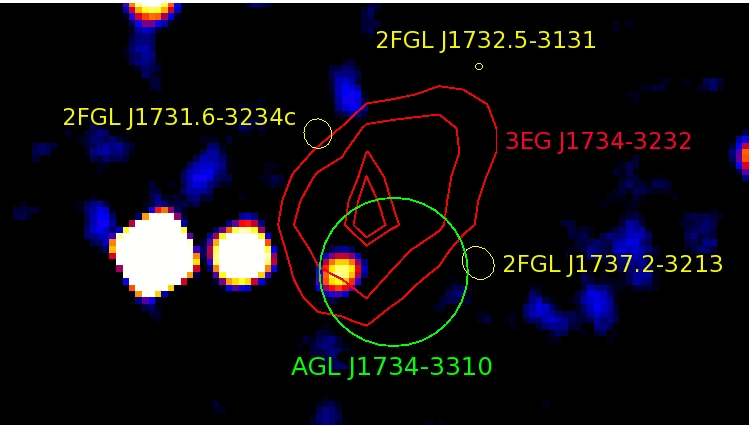}
\caption{The INTEGRAL/IBIS mosaic significance map (18--60 keV, $\sim$ 10 Ms exposure time) of the sky region including IGR J17354$-$3255. 
The other two bright sources detected in the field are the LMXBs GX 354$-$0 and 4U 1730$-$335.
The green error circle represents the MeV-GeV source AGL J1734$-$3310 and the red contours (from 50\%  to 99\%) refer to 
3EG J1734$-$3232. The persistent gamma-ray sources detected by Fermi/LAT are indicated by means of yellow gamma-ray error circles} 
\end{center}
\end{figure*}

\section{The SFXT IGR J17354$-$3255 as test case of gamma-ray emitter}
Interestingly,  the SFXT  IGR~J17354$-$3255 is located in the sky region of two unidentified gamma-ray  sources:  
AGL J1734$-$3310 and 3EG J1734$-$3232 (see Fig. 4):

\begin{itemize}
\item AGL J1734$-$3310 was discovered by the AGILE gamma--ray satellite on 2009 April 14 during a flare  lasting 
only 1 day and detected with a significance of  4.8$\sigma$ at  E$>$100 MeV \cite{bulgarelli}. 
After its discovery, extensive searches for further flaring  gamma-ray emission have been carried  out by the AGILE team \cite{bulga2}.  As a result, several  additional gamma-ray  flares have been discovered in the AGILE data archive 2007--2009. They have a similar duration (about 1 day) and significance detection in the range (3--5)$\sigma$. This clearly shows that AGL J1734$-$3310 is a recurrent transient gamma-ray  source.  The significance of the sum of all the flares detected by AGILE is 7.3$\sigma$ with a 95\%  statistical 
and systematic positional error radius of 0.46 degrees. 
Fig. 4 shows the 18--60 keV INTEGRAL/IBIS  significance mosaic map ($\sim$ 10 Ms exposure) of the sky region 
surrounding IGR J17354$-$3255 with superimposed the positional uncertainty of AGL J1734$-$3310 (green circle). 
Clearly, IGR J17354$-$3255 is the only hard X-ray source  detected inside the AGILE error circle, 
the same holds in the softer X-ray band (3--10 keV) from an INTEGRAL/JEM--X deep mosaic ($\sim$ 700 ks exposure). 
We took into account the possibility of a chance coincidence and to this aim we calculated the probability of finding a supergiant HMXB, such as IGR J17354$-$3255, inside the AGILE error circle by chance. Given the number of supergiant HMXBs detected by IBIS within the Galactic plane \cite{bird}, defined here as restricted to a latitude range of $\pm$5$^{\circ}$, we estimated a probability of  $\sim$1$\%$.

\item 3EG J1734$-$3232 is a still unidentified gamma-ray source listed in the  third EGRET catalog with average flux 
of (40$\pm$6.7)$\times$10$^{-8}$ photons cm$^{-2}$ s$^{-1}$ (E $>$100 MeV) and significance detection of 6.2$\sigma$  \cite{hart}. 
It is also designated as confused source, which means that it may have significant uncertainties due to the overlapping PSFs. It is in fact
possible that 3EG J1734$-$3232 is  the blend of more than one gamma-ray source, this is particularly evident thanks to recent observations performed by AGILE and Fermi whose excellent angular resolution allowed to pinpoint the two gamma-ray sources likely responsible for the entire emission from 3EG J1734$-$3232 (see Fig. 4): AGL J1734$-$3310 (discussed above) and 2FGL J1732.5$-$3131. 
The latter is a firmly identified gamma-ray pulsar \cite{nolan} with average flux  $\sim$ 20$\times$10$^{-8}$ photons cm$^{-2}$ s$^{-1}$ (100 MeV--100 GeV). Such hypothesis is strongly supported by i) the blending  of 3EG J1734$-$3232 elongated towards the direction of both AGL J1735$-$3258 and 2FGL J1732.5$-$313 ii) the  average EGRET gamma-ray flux compatible with the sum of  the gamma-ray fluxes from 
the AGILE and Fermi sources  (i.e. $\sim$45$\times$10$^{-8}$ photons cm$^{-2}$ s$^{-1}$ at E $>$100 MeV). In addition,  3EG J1734$-$3232 is likely variable as  suggested by the value of its  variability index I \cite{cina}. The spatial match, the possible variability, the gamma-ray flux values, all strongly suggest that at least a fraction of the entire gamma-ray emission from 3EG J1734$-$3232 might well be associated with AGL J1735$-$3258. The remaining part is very likely coming from 2FGL J1732.5$-$3131.

\end{itemize}

For the sake  of completeness, we note that in the surroundings of  AGL J1734$-$3310 and 3EG J1734$-$3232
 there are  two other  persistent gamma-ray  sources \cite{nolan} as detected by Fermi (see Fig. 4):   i) 2FGL J1737.2$-$3213 is an unidentified gamma-ray source with average flux $\sim$ 11$\times$10$^{-8}$ photons  cm$^{-2}$ s$^{-1}$ (300 MeV--100 GeV), it is not variable and  it is likely associated with a supernova remnant or pulsar wind nebula  ii) 2FGL J1731.6$-$3234c is still unidentified  with average flux of $\sim$ 6$\times$10$^{-8}$ photons  cm$^{-2}$ s$^{-1}$  (300 MeV--100 GeV), however it is found in a region with possibly incorrected diffuse emission.  As such, its position and even existence may not be reliable, i.e.  it could be a fake source potentially confused with interstellar emission.  We note that none of the above three Fermi gamma-ray sources is spatially associated with AGL J1734$-$3310  and/or  IGR J17354$-$3255.  Moreover,  despite 2FGL J1737.2$-$3213 and 2FGL J1731.6$-$3234c are in the surroundings of 3EG J1734$-$3232,  they  likely give no significant contribution to the flux measured by EGRET since both are very weak gamma-ray sources. 

\section{Conclusions}
Based on spatial correlation as well as on a  flaring nature  on similar short timescales (although not simultaneous yet),  we propose the SFXT IGR~J17354$-$3255 as the best candidate counterpart of AGL~J1734$-$3310, to date. 
Although such proposed association is merely based on intriguing hints, it represents an important first step towards 
obtaining a reliable test case on which to concentrate further efforts to obtain 
quantitative proofs for a real physical association. In this respect, further AGILE, Fermi  and INTEGRAL studies of IGR J17354$-$3255/AGL J1734$-$3310  are under way.  If fully confirmed, the implications of SFXTs producing 
gamma-ray emission are huge, both theoretically and observationally since  i) it would open the study to an unexplored energy window, ii) it would  allow a deep inspection of the extreme physical mechanisms able to accelerate particles up to relativistic energies in HMXBs, iii)  it would add  a further extreme characteristic  to the already extreme class of SFXTs.

\section{acknowledgments}
V. Sguera is very grateful to Andrea Bulgarelli and the AGILE team for sharing the results on AGL~J1734$-$3310 before publication





\begin{thebibliography}{00}

 
 \bibitem{hays}	
Hays et al. 2009, AAS Meeting 213, 612.04
 
 \bibitem{bulgarelli} 
 Bulgarelli et al. 2009, ATel 2017
 
 \bibitem{sguera1} 
 Sguera, V.,Romero,G.E.,Bazzano, A.,et al. 2009, ApJ, 697,1194 
 
 \bibitem{sguera2} 
 Sguera, V., Drave, S. P., Bird, A. J., et al. 2011, MNRAS, 417, 573  
 
 \bibitem{sguera3} 
 Sguera, V. 2009, arXiv 0902.0245, proceedings of the 7th INTEGRAL Workshop, PoS Integral08:082,2008 
 
 \bibitem{sguera4} 
 Sguera, V., Barlow, E. J., Bird, A. J., et al. 2005, A\&A, 444, 221
 
\bibitem{sguera5} 
Sguera, V., Bazzano, A., Bird, A. J., et al. 2006, ApJ, 646, 452 
 
 \bibitem{negueruela}
 Negueruela, I.; Smith, D. M.; Reig, P.;, et al. 2006, ESA SP, 604, 165 
 
 
 \bibitem{sguera6} 
 Sguera, V., Hill, A. B.; Bird, A. J., et al. 2007, A\&A, 467, 249 
 
 \bibitem{clark} 
 Clark, D.J., Sguera, V., Bird, A.J. et al. 2010, MNRAS, 406L,75  
 
 \bibitem{wz} 
 Walter, R. \& Zurita Heras, J., 2007, A\&A, 476, 335 
 
 \bibitem{grebenev}	
 Grebenev 20120, arXiv 1004.0293, Proceedings of the Workshop "The Extreme sky: Sampling the Universe above 10 keV", PoS, 96, 60 
 
 \bibitem{sidoli1} 
 Sidoli, L., 2009, AdSpR, 43, 1464  
 
 \bibitem{mirabel}	
 Mirabel 2012, Science,  335, 175  
 
 \bibitem{tavani} 
 Tavani, M., Bulgarelli, A., Piano, G., et al. 2009, Nature, 462, 620 
 
 \bibitem{abdo} 
Abdo  et al. 2009, Science, 326, 1512

\bibitem{sabatini}
Sabatini S. et al., 2010, ApJ, 712, L10 

\bibitem{k1} 
Kuulkers, E. et al., 2006, ATel 874 

\bibitem{k2} 
Kuulkers, E.; Shaw, S. E.; Paizis, A.; et al. 2007, A\&A, 466, 595 

\bibitem{dai} 
D'Ai et al. 2011, A\&A, 529, 30

\bibitem{ver} 
Vercellone, S., D'Ammando, F.,  Striani, E., et al. 2009, ATel 2019

\bibitem{tom} 
Tomsick, J.A., Chaty, S., Rodriguez, J. et al. 2009, ApJ, 701, 811

\bibitem{boz} 
Bozzo, E.; Pavan, L.; Ferrigno, C., et al. 2012, A\&A, 544, 118

\bibitem{coleiro} 
Coleiro et al. 2012, poster presented at the 9th INTEGRAL conference held in Paris, 15-19 October 2012

\bibitem{bulga2}
Bulgarelli et al. in preparation 

\bibitem{bird} 
Bird, A. J., Bazzano, A., Bassani, L., et al. 2010, ApJS, 186, 1

\bibitem{hart}	
Hartman, R. C. et al. 1999, ApJS, 123, 79

\bibitem{nolan} 
Nolan et a. 2012, ApJs, 199, 2, 31 

\bibitem{cina}
Han \& Zhang 2005, Chinese Journal of Astronomy and Astrophysics, 5, 3, 256 


 \end{thebibliography}



\end{document}